\documentclass[11pt]{article}

\usepackage[utf8]{inputenc}
\usepackage[T1]{fontenc}
\usepackage[english]{babel}
\usepackage{amsmath,amssymb,amsthm}
\usepackage[round,authoryear]{natbib}
\usepackage{hyperref}
\usepackage{doi}
\usepackage{geometry}
\usepackage{setspace}
\hypersetup{
  colorlinks=true,
  linkcolor=black,
  urlcolor=blue,
  citecolor=black
}

\title{Property, Interest, and Money:\\
  Is Heinsohn and Steiger's Property Premium\\
  a Determinant of Interest?}
\author{Eric Hillebrand\thanks{Department of Economics and Business Economics, Aarhus University, Denmark. Email: \href{mailto:ehillebrand@econ.au.dk}{ehillebrand@econ.au.dk}.}}
\date{\today}

\begin{document}
\maketitle

\begin{abstract}
\noindent Heinsohn and Steiger's \emph{Eigentum, Zins und Geld} (1996) proposes the property premium as the foundational determinant of interest, replacing time preference. This paper examines whether the replacement succeeds. It does not. The two arguments against time preference, the savings-inelasticity claim after Hahn and the portfolio-shift claim after Keynes, both fail on standard microeconomic grounds. With time preference intact, the property premium sits within the standard decomposition of the interest rate. In ordinary collateralized credit it coincides with the risk premium. Only when the lender is a money-issuing bank with a real redemption obligation does a third term enter the decomposition that standard asset-pricing theory does not articulate. That third term is Heinsohn and Steiger's genuine contribution. The paper discusses its apparent disappearance or disguised operation after 2008, and the circularity of a property anchor measured in money.
\end{abstract}

\section{Introduction}

The argument of \emph{Eigentum, Zins und Geld} rests on a distinction that is standard in law but rarely emphasized in standard economics: the distinction between property (\emph{Eigentum}) and possession (\emph{Besitz}). Property is a bundle of legal titles: intangible, contractually disposable rights that exist only in societies that have created a legal order recognizing them. Possession is the physical use of things and exists in every human society, including those without property institutions. The distinction is not semantic; it is the foundation of the entire theory. In tribal and feudal orders there is possession but no property, and in those orders there is lending of goods but no interest \citep[pp.~463--464]{HeinsohnSteiger2004}.\footnote{All page references to \citet{HeinsohnSteiger2004} are to the third edition, the most recent German edition and the one cited throughout this paper. Concordance with the first edition (Reinbek bei Hamburg: Rowohlt, 1996) and the English version edited by Frank Decker (London: Routledge, 2013) can be supplied on request. For the detailed treatment of the \emph{Eigentum/Besitz} distinction, see Part B, \emph{Das Kapitel vom Eigentum}.}

The core theoretical construction follows from this distinction. When property exists, holding it unencumbered yields an immaterial return: the property premium (\emph{Eigentumspr\"amie}). This premium consists not of any physical product but of the economic security that comes from owning disposable, pledgeable, encumberable rights: the capacity to back the issuance of money, to serve as collateral for obtaining credit, and to be sold. It is a yield on a legal position, not on a physical asset \citep[pp.~463--464; the core positive construction is in Part C, section 3, \emph{Die Eigentumstheorie des Zinses}, pp.~173ff.]{HeinsohnSteiger2004}.

Heinsohn and Steiger argue that when a creditor enters a credit contract, they encumber their property, block it, and thereby lose their property premium for the duration of the contract. The debtor must compensate them for this loss. That compensation is interest. In the authors' formulation, the claim to compensation for the foregone property premium manifests itself as interest \citep[p.~464, thesis 8]{HeinsohnSteiger2004}. Interest is not the price of deferred consumption, not the reward for abstinence, and not the return on physical capital. It is the compensation for the temporary loss of the immaterial yield on unencumbered property.

The claim is explicitly one of replacement, not complement. The neoclassical explanation of interest through time preference, going back through \citet{Fisher1930} to \citeauthor{BohmBawerk1889}'s three grounds \citep[pp.~59--60]{HeinsohnSteiger2004}, is rejected. So is Keynes's explanation through liquidity preference, which the authors credit with having correctly displaced the first, neoclassical time preference but then criticize for falling into a ``second psychological time preference'' without reaching the property premium \citep[pp.~194--201]{HeinsohnSteiger2004}. The property premium is to stand where all three, time preference, marginal productivity of capital, and liquidity preference, previously stood. Property premium and interest are characterized as the primordial value and primordial price of the property-based economic order \citep[p.~340]{HeinsohnSteiger2004}.

The reception of \emph{Eigentum, Zins und Geld} in peer-reviewed journals has been thin given the ambition of the project. The early book reviews by \citet{Graziani1997} and \citet{Backhaus2000} welcomed the theory as an original contribution while flagging overclaims of originality relative to existing property rights traditions. The English edition of a generation later drew reviews from \citet{Luther2015} and \citet{Sauer2015}, mostly descriptive rather than argumentative. \citet{Riese1985}, writing in \emph{Leviathan} a decade before the first edition of the book appeared, gave the first substantive journal engagement with the program. Beyond reviews, \citet{Steiger2006}, writing in the \emph{Journal of Economic Issues}, is the canonical English-language self-statement of property economics.\footnote{In a parallel collaboration with Hans-Joachim Stadermann, Otto Steiger developed a related framework that places the emphasis on the obligation side of the contract rather than on the property side. We do not pursue that line further here; it is set out in \citet{StadermannSteiger2001} and \citet{Stadermann2006}.} \citet{Decker2015, Decker2017} continue the program in the same journal and in \emph{Economic Affairs} respectively. \citet{GerberSteppacher2017}, in \emph{Anthropological Theory}, extend the theory toward possession-based economies. The broader property rights discussion in \citet{Hodgson2015} engages the tradition that Heinsohn and Steiger position themselves against rather than the property premium itself. The bulk of the substantive debate has taken place in Metropolis-Verlag edited volumes \citep{Steiger2008ed,BeaufortDecker2016}. Between the review-level engagements and the in-house continuation, the middle ground of critical engagement has been sparsely populated.

This paper examines whether the replacement of time preference with the property premium succeeds. Specifically, it asks whether Heinsohn and Steiger's arguments against time preference are valid, and what follows for the status of the property premium if they are not. The distinction between property and possession is the book's foundational move, and I take it as given rather than critique it here. I also set aside the other dimensions of the work, such as the historical claims, the origin of money in the credit contract, and the critique of the barter narrative.

The paper is organized as follows. Section~\ref{sec:savings} examines the first of Heinsohn and Steiger's two attacks on time preference: the apparent inelasticity of savings to the interest rate. Section~\ref{sec:keynes} turns to the second attack, which locates the determination of the interest rate not at the consumption decision margin but at the portfolio allocation margin, largely following Keynes. Section~\ref{sec:decomp} asks where the property premium belongs in standard asset-pricing decompositions of the interest rate. Section~\ref{sec:bank} treats the special case of a money-issuing bank that creates redeemable money in a credit contract: the paradigmatic example and original contribution of Heinsohn and Steiger. Section~\ref{sec:after2008} considers whether the receding ability of modern central banks to revert to redeemable issuance is pertinent at all, and if it is, where the ramifications may become visible. Section~\ref{sec:circle} concludes by asking whether property evaluated in monetary terms can serve as a non-circular anchor for money issuance.

\section{The Savings-Inelasticity Argument}
\label{sec:savings}

The first of Heinsohn and Steiger's two attacks on time preference is empirical. In Part A of \emph{Eigentum, Zins und Geld} \citep[pp.~56--57]{HeinsohnSteiger2004} the authors set up a neoclassical Robinson who, faced with a higher interest rate, ought to defer more consumption. From this they extract the prediction that aggregate savings should rise with the rate of interest. They then dispatch the prediction with a single citation: Albert Hahn, the Frankfurt bank director, observing in 1920 from his counter at the bank that nothing influences savings less than the level of interest rates (\emph{``dass kein Moment die Spart\"atigkeit weniger beeinflusst als die Zinsh\"ohe''}) \citep[p.~154]{Hahn1920}. The neoclassical assumption that savings respond to the interest rate is, on this basis, dismissed as scientific folklore \citep[p.~57]{HeinsohnSteiger2004}. The promise is that the chapter on interest will return to the matter and complete the replacement.

Two things must be said about this argument. The first is empirical: the citation is thin. A single pre-econometric observation by a banker, however sharp, is not evidence in any modern sense. The actual empirical literature on the interest elasticity of savings is mixed and context-dependent, ranging from the substantial positive estimates of \citet{Boskin1978} and the even larger theoretical predictions of \citet{Summers1981}, through the near-zero elasticity of intertemporal substitution found by \citet{Hall1988}, to the comprehensive surveys of \citet{BrowningLusardi1996} and \citet{AttanasioWeber2010} that confirm the evidence has not converged, and is in any case not the clean knockout the argument requires.

The second is more fundamental. Even if Hahn were exactly right, and the interest elasticity of aggregate savings were precisely zero, this would not eliminate time preference. The inference that Heinsohn and Steiger draw, \emph{no response of savings to interest, therefore no time preference}, is a non sequitur. The reason is the standard Slutsky decomposition of any price change into a substitution effect and an income effect, applied here to the price of present consumption (which is what the interest rate is). Both effects are present, they generally point in opposite directions, and their sum can be of either sign. For a saver they can exactly cancel.

The textbook treatment is in \citet[chaps.~8--10]{Varian2010}; for a treatment at the graduate level, see \citet[chap.~3, sections 3.D--3.G]{MWG1995}. The formal derivation for the intertemporal case is collected in Appendix~\ref{app:slutsky}. Two facts from that derivation matter for the argument here:

\begin{enumerate}
\item \emph{The substitution effect always raises savings.} When the interest rate rises, present consumption becomes relatively more expensive; the agent substitutes toward future consumption, which means saving more.
\item \emph{The income effect for a saver is opposite in sign.} A higher interest rate makes a net saver wealthier in lifetime terms. With present consumption normal, the agent wants more of it, which means saving less.
\end{enumerate}

The net effect on savings is the algebraic sum. It is theoretically ambiguous and empirically an open question, depending on preferences, the intertemporal elasticity of substitution, and the share of savers vs.\ borrowers in the population. The clean prediction Heinsohn and Steiger attribute to neoclassical theory is not, in fact, a neoclassical prediction. \citet{Fisher1930}, the foundational neoclassical reference for the theory of interest, does not make it.

The cancellation can be made fully explicit. With logarithmic utility, the canonical case with intertemporal elasticity of substitution equal to one, and a saver whose entire income is concentrated in the first period, the two effects exactly offset. Period-1 consumption (and therefore savings) is independent of the interest rate. Hahn's observation is precisely what this special case predicts. It is \emph{consistent with} a positive rate of time preference, not evidence against one.

The deeper point is that the rate of time preference $\rho$ and the intertemporal elasticity of substitution are \emph{independent} parameters of the agent's preferences. The first governs the level of savings; the second governs their responsiveness to relative price changes. Zero responsiveness implies nothing about the level. To eliminate $\rho$ from economic theory one would need to show that the entire joint parameter space, every combination of $\rho$ and the IES consistent with positive time preference, is incompatible with the data. No econometric work has ever come close to such a claim, and certainly Hahn's banker's anecdote does not.

The first argument therefore fails on both fronts. The empirical premise is unsupported, and even if granted it does not deliver the conclusion. Time preference survives intact.

\section{The Keynes Portfolio-Shift Argument}
\label{sec:keynes}

The second attack is theoretical and is laid out in Part C, section 4a \citep[\emph{Liquidit\"atspr\"aferenz, Geldnachfrage und Zins bei Keynes}, pp.~194--201]{HeinsohnSteiger2004}. Here Heinsohn and Steiger credit Keynes with having convincingly refuted the neoclassical time-preference theory of interest. The structure of the alleged refutation is as follows. The agent faces \emph{two} decisions, not one. The first is the consumption-saving decision: how much of period-1 income to consume now and how much to defer. The second is the portfolio decision: in what \emph{form} to hold the deferred amount: as cash, which earns no interest, or as a bond, which does. Keynes's point, as the authors transmit it, is that interest is paid only at the second margin. An agent who saves by holding cash earns nothing. Interest is the price of parting with liquidity, not the price of saving as such. Heinsohn and Steiger conclude that time preference has thereby been leveraged out of the theory of interest.

Keynes's observation is correct and important, and it deserves the place Heinsohn and Steiger give it. Interest does enter at the portfolio margin. An agent can save without lending, and saving as such carries no remuneration. What does \emph{not} follow is that time preference has been eliminated. The inference confuses two different questions:

\begin{enumerate}
\item \emph{Where in the market is the interest rate determined?} Keynes's answer: at the portfolio allocation margin, where the price of illiquidity is set by the public's collective preference for liquidity.
\item \emph{What governs how much an individual agent saves at a given interest rate?} This is a preference question, answered by the agent's first-order condition for the consumption-saving choice, and time preference lives inside that condition.
\end{enumerate}

The two questions have different answers, and answering the first without invoking $\rho$ does not remove $\rho$ from the second. To see this it is enough to write down the two-stage problem in the simplest form and read off the optimality conditions. The full derivation is in Appendix~\ref{app:euler}; the result is that for any $r > 0$ the agent strictly prefers bonds to cash (cash is dominated), and the consumption-saving margin is then governed by the standard Euler equation:
\[
u'(c_1) \;=\; \beta(1+r)\,u'(c_2), \qquad \beta = \frac{1}{1+\rho}.
\]
The discount factor $\beta$, and therefore the time preference rate $\rho$, sits in the very condition that determines how much the agent saves at the prevailing interest rate. Keynes's relocation of \emph{where interest enters the market} does nothing to remove $\rho$ from \emph{how the agent allocates consumption across time}.

The previous section showed that $\rho$ cannot be inferred from the slope $\partial s/\partial r$; the present section shows that $\rho$ governs the level of $s$ itself, via the Euler equation that survives the relocation of the interest rate to the portfolio margin.

The further claim Heinsohn and Steiger make in this section, that Keynes himself, having rejected the first time preference, falls into a \emph{``second psychological time preference''} in the form of liquidity preference, which he then, in Heinsohn and Steiger's view, cannot ground because he lacks the property premium, is a separate issue and is consistent with the standard reading of Keynes by his interpreters \citep{Robertson1936,Hicks1937,Hicks1939}. It does not bear on the question of whether the \emph{first} time preference, the preference parameter governing intertemporal substitution, has been eliminated. It has not.

A simple thought experiment confirms the conclusion. Consider a credit contract with zero default risk: a perfectly safe borrower, no possibility of loss, no collateral exposure. The property premium, which, as developed in the next section, is in this general case a risk premium in new vocabulary, is zero. Yet the lender will still demand a positive interest rate, because consumption today is worth more to them than the same consumption tomorrow. That residual rate is what time preference names. No alternative concept in \emph{Eigentum, Zins und Geld} explains it.

The two-step replacement of time preference therefore fails at both steps. The empirical step fails because the inference from zero elasticity to no time preference is invalid. The theoretical step fails because the relocation of the market for interest does not touch the preference parameter that governs the agent's saving choice. With time preference still in place, the property premium cannot occupy the foundational position the authors design for it. Where it can stand, and where it makes a contribution that no standard concept covers, is the subject of Sections \ref{sec:decomp} and~\ref{sec:bank}.

\section{The Standard Decomposition and the Property Premium in the General Case}
\label{sec:decomp}

With time preference not eliminated, the familiar two-term decomposition of the market interest rate remains the reference framework against which any alternative theory must be read:
\[
\tilde{r} \;=\; r_f \;+\; (\tilde{r} - r_f),
\]
where $r_f$ is the risk-free rate and $\tilde{r} - r_f$ is the risk premium compensating the lender for exposure to default. This is the workhorse of modern asset-pricing theory. The question this section addresses is where the property premium sits within it. In the general case of collateralized credit between private agents, the Heinsohn/Steiger paradigm situation outside the special case of money-issuing banks, which is reserved for Section~\ref{sec:bank}, the answer is that the property premium coincides with the risk premium. The institutional vocabulary is richer; the analytical content is the same.

\paragraph{The first term.} The risk-free rate is the rate a lender demands on a loan that cannot default. Appendix~\ref{app:euler} established its identity with the rate of time preference: when consumption is stationary the Euler equation collapses to $1 + r_f = 1/\beta = 1 + \rho$, so $r_f = \rho$. With $\rho > 0$, a perfectly safe loan commands a strictly positive interest rate purely because present consumption is preferred to future consumption. This is the residual rate that the arguments in \emph{Eigentum, Zins und Geld} against time preference do not reach and that no alternative concept in the book explains.

\paragraph{The second term.} The risk premium is the compensation for bearing default risk. The derivation is given in full in Appendix~\ref{app:risk}. For a risk-neutral lender the spread is approximately
\[
\tilde{r} - r_f \;\approx\; \pi\,\lambda,
\]
where $\pi$ is the probability of default and $\lambda \in [0,1]$ is the loss given default, that is, the expected loss per unit lent. For a risk-averse lender a further variance-related term is added. The qualitative structure is robust to these refinements: $\tilde{r} - r_f$ is the market price of the lender's default exposure.

\paragraph{Locating the property premium.} Consider now the paradigm Heinsohn/Steiger contract: a creditor lends to a debtor who pledges property as collateral. With probability $\pi$ the debtor defaults and the creditor loses a fraction $\lambda$ of the pledged property. The compensation the creditor rationally demands for accepting that \emph{their exposure to the loss of property} runs for the duration of the contract is $\pi \lambda$ plus a risk-aversion term. This is exactly the risk premium derived in Appendix~\ref{app:risk}.

Heinsohn and Steiger's framing gives the mechanism a distinct institutional reading. The lender does not merely bear an expected monetary loss; they give up the immaterial security yield on unencumbered property: the freedom to dispose of it, pledge it, or sell it for the duration of the contract. That is the property premium as the authors describe it \citep[pp.~173ff.; thesis 8, p.~464]{HeinsohnSteiger2004}. The description is institutionally illuminating and the language of \emph{property at risk} reads further into the legal and historical preconditions of credit than the language of \emph{value at risk} does. But for ordinary collateralized credit between individuals the object the two descriptions refer to is the same object. The premium is compensation for the possibility of loss, discounted by its probability and weighted by the lender's attitude to risk. Whether the object lost is named \emph{property} or \emph{value}, the expected-loss-plus-risk-aversion formula is the same, and the market-clearing premium is the same.

In the general case, then, the property premium does not stand alongside time preference as a separate foundational term. It is the standard risk premium rendered in the vocabulary of property. The replacement of time preference fails, Sections 2 and 3, and the supplementary concept the authors introduce reduces, in this general case, to a concept the theory already has.

This verdict is not the end of the story. There is one institutional configuration in which the argument of this section breaks down: the case where the lender is not a private individual, or a modern bank, but a money-issuing bank whose loan creates circulating notes redeemable against the bank's own property (\emph{Zettelbank}, in German). In that configuration, and, as will be argued, \emph{only} in that configuration, a third term enters the decomposition that standard asset pricing theory does not articulate. Isolating and interpreting that term is the business of Section~\ref{sec:bank}.

\section{The Money-Issuing Bank and the Third Term}
\label{sec:bank}

The configuration in which the property premium does \emph{not} reduce to a standard risk premium, and where, accordingly, the Heinsohn/Steiger framework makes a contribution that asset-pricing theory does not articulate, is the one the authors develop at greatest length in Part D: the money-issuing bank under competitive issuance with a real redemption obligation. In that configuration a third term enters the decomposition. Isolating it and asking what it can and cannot be used to identify is the business of this section.

\paragraph{The mechanism.} When a bank makes a loan by issuing its own circulating notes, the notes are liabilities of the bank redeemable on demand against its property: gold, notes of other banks, or some other pledgeable asset. The loan runs to maturity; the notes do not. Once in circulation they may be presented for redemption by any third-party holder, anonymous to the bank, at any moment before the loan matures. The bank must therefore keep its balance sheet in a condition to honour redemption throughout the life of the loan. It must, as Heinsohn and Steiger put it, stand ready with unencumbered property. For the duration of the loan it cannot dispose of that property as it otherwise would, and forgoing this disposition is a real cost.

The mechanism is not a gloss the authors introduced for effect. They describe it precisely in Part D in connection with the central bank's discounting of a bill of exchange \citep[pp.~291--292]{HeinsohnSteiger2004}. The discount, they argue, is not a tool of monetary control but compensation for exactly this encumbrance: the bank takes in a claim collectible only in the future and pays out its own liabilities redeemable immediately, blocking its property over the interval in between. The bill-creditors do not bear this temporary loss of the property premium on the bank's property. The bank therefore has itself compensated for it through interest, which it deducts from the bill-creditor as the discount \citep[p.~291]{HeinsohnSteiger2004}. The bank's own property premium, temporarily forgone, is the object the discount compensates.

\paragraph{The third term.} The cost just described is not time preference: it does not arise from any intertemporal substitution in consumption and is not $\rho$. It is not credit risk on the loan: the bank's exposure to redemption runs independently of whether the borrower defaults, and exists even when $\pi = 0$. It is a distinct cost, arising specifically from the mechanics of money creation through credit. Writing the interest rate charged by a money-issuing bank as a decomposition of its three components gives
\[
\tilde{r}_{\text{Zettelbank}} \;=\; \underbrace{r_f}_{\substack{\text{time preference}\\(\approx \rho)}} \;+\; \underbrace{(\tilde{r} - r_f)_{\text{default}}}_{\text{standard risk premium}} \;+\; \underbrace{\phi}_{\substack{\text{property encumbrance}\\ \text{premium}}}.
\]
The first two terms are the components of Section~\ref{sec:decomp}. The third, $\phi$, captures the shadow cost per unit lent of maintaining the bank's redemption capacity. A formal statement is given in Appendix~\ref{app:bank}; the content that matters for the argument here is the decomposition itself and its behaviour in the two limiting cases:

\begin{itemize}
\item For lending between individuals with no money creation, $\phi = 0$ and the decomposition collapses to the two-term form of Section~\ref{sec:decomp}.
\item For a money-issuing bank with a binding redemption obligation, $\phi > 0$ and persists even when the loan is perfectly safe.
\end{itemize}

The third term is the component that standard asset pricing theory does not name. It arises because the institutional configuration, competitive issuance plus real redemption, makes the \emph{lender's} property, not the borrower's, the object at risk. In ordinary credit this effect is negligible. Where the lender creates money, it becomes systemic. This is, in our reading, the deepest insight of \emph{Eigentum, Zins und Geld}, and it is the component of the framework that cannot be reproduced in any of the existing alternatives.

\paragraph{Verdict on the replacement claim.} The three-term decomposition also makes the limits of the Heinsohn/Steiger project exact. Their claim that the property premium replaces time preference fails: $r_f \approx \rho$ persists as a separate term that the arguments in the book do not reach. Their claim that property at risk is the key to understanding interest is correct but, for ordinary credit, unsurprising: it is a renaming of the standard risk premium. Their genuine contribution is the third term $\phi$, which captures a real economic cost that standard theory does not articulate because it arises only from a specific institutional configuration that standard theory has left to economic history. The framework does not replace the existing theory of interest; it augments it with a component standard theory lacks.

\paragraph{Identification: why modern data cannot separate $\phi$.} This is the point at which the econometric question must be raised, because it determines what can and cannot be done with the three-term decomposition empirically. In modern banking data the spread of a bank's lending rate over the risk-free rate bundles default risk, liquidity risk, operational costs, capital requirements, and $\phi$ into a single observed quantity. For central banks with a monopoly on issuance and no real redemption obligation, the redemption window is institutionally closed, and $\phi$ has been engineered out. No one can walk into the European Central Bank (ECB) with a banknote and claim property. For commercial banks, deposit-issuance superficially resembles the historical note-issuance case, but deposit insurance, the lender-of-last-resort function, and central bank liquidity backstops absorb the redemption pressure that would otherwise price $\phi$. In each case $\phi$ is either zero or observationally equivalent to the standard risk premium. Contemporary time-series on bank lending rates therefore cannot be used to identify $\phi$. The attempt collapses into collinearity.

\paragraph{The case for cliometrics.} Identification requires variation that contemporary data do not provide: an institutional regime in which redemption pressure varies independently of default risk, so that movements in $\phi$ can be separated from movements in the credit risk premium. This variation is available historically. Free-banking episodes with genuine redemption offer the required cross-sectional and episode-level variation: the Scottish system c.\ 1716--1845 \citep{White1995}, the US free-banking era 1837--1863 \citep{RolnickWeber1983,RolnickWeber1984}, the early Bank of Amsterdam \citep{QuinnRoberds2014}, the Italian city-state banks treated in Part D of \emph{Eigentum, Zins und Geld}. The relevant cliometric design would exploit differences across banks, or across regimes, in redemption frequency and the cost of maintaining redemption capacity, holding default fundamentals fixed. Heinsohn and Steiger narrated these regimes extensively but did not exploit them econometrically. Neither did they collaborate with anyone who would have. This is, we suspect, the most productive direction in which their framework can still be taken: a cliometric programme aimed at isolating $\phi$ in the regimes where it is identifiable, to establish its empirical magnitude and its sensitivity to institutional detail.

The question, then, is whether modern fiat monopoly has in fact eliminated the mechanism that generated the property premium or whether it has merely suppressed it and it still operates in disguised form. I offer some thoughts on this in Section~\ref{sec:after2008}.

\section{Elimination or Disguised Operation}
\label{sec:after2008}

Section~\ref{sec:bank} closed on the question of whether modern fiat monopoly has eliminated the mechanism that generated the property premium or has merely suppressed it, and signalled that I would offer some thoughts here rather than attempt to resolve it. The section proceeds in four steps. The first establishes the empirical record against which any answer must be read. The second recovers the Heinsohn/Steiger normative tenet on central bank asset backing and proposes a reading of it as maintenance of an unexercised reversion option. The third and fourth present, in turn, the case that the property premium has been eliminated and the case that it still operates in disguised form. Section~\ref{sec:circle} then steps outside the elimination/disguised-operation frame to consider a puzzle that complicates both readings and that the paper can only flag, not resolve.

\subsection{The Empirical Record}

The size of the balance sheets is the first-order fact. On the eve of the Global Financial Crisis (GFC) in August 2007 the Federal Reserve held total assets of roughly \$870 billion; a decade and a half of quantitative easing (QE) and other unconventional monetary policy later, the balance sheet peaked at close to \$8.9 trillion at the end of the first quarter of 2022, roughly a tenfold expansion. It stood at approximately \$6.5 trillion at the end of 2025 after three and a half years of tightening, meaning only about half of the pandemic-era expansion has been reversed and the post-2008 plateau has not been re-approached. The Eurosystem traces the same arc on a similar scale. The consolidated balance sheet of the ECB and the national central banks stood at roughly \texteuro 1.5 trillion at the end of 2007 and grew to a peak of \texteuro 8.56 trillion at the end of 2021; it had fallen to \texteuro 7.95 trillion by end-2022 and to comparable levels since. Both arcs are punctuated by the same three breakpoints: 2008 initiated the regime change, the Covid response of 2020--21 roughly doubled the already-expanded balance sheets, and the Ukraine-era inflation shock of 2022 triggered the tightening that has since only partially reversed the pandemic expansion.\footnote{Sources for the balance-sheet totals. Fed weekly total assets series (H.4.1 / WALCL): \url{https://fred.stlouisfed.org/series/WALCL}. Congressional Research Service overview of the Federal Reserve's balance sheet since 2007: \url{https://www.congress.gov/crs-product/IF12147}. Eurosystem annual consolidated balance sheet 1999--2025: \url{https://www.ecb.europa.eu/press/annual-reports-financial-statements/annual/balance/html/index.en.html}. ECB commentary on the 2021 balance sheet: \url{https://www.ecb.europa.eu/press/annual-reports-financial-statements/annual/balance/html/ecb.eurosystembalancesheet202206_commentary~fa1d143aa2.en.html}. ECB commentary on the 2022 balance sheet: \url{https://www.ecb.europa.eu/press/annual-reports-financial-statements/annual/balance/html/ecb.eurosystembalancesheet202306_commentary~786c61e933.en.html}.}

More instructive, for the question this paper is pressing, is the maturity structure and composition of the asset side. Pre-2008, the Fed's System Open Market Account (SOMA) was weighted toward short-dated paper: the portfolio was dominated by Treasury bills maturing within a year, and at the end of 2006 less than 20\% of the Fed's Treasury holdings matured more than five years out. The Desk purchased securities across the yield curve specifically to avoid distorting it, and the portfolio held essentially no mortgage-backed securities (MBS) at all. This posture was close, though not identical, to the Bundesbank's bill-of-exchange (\emph{Handelswechsel}) practice: short-dated, private-sector, market-priced, renewable at high frequency, with the composition of the portfolio continuously validated by the market through the rolling of maturities. Post-2008 that posture has been progressively abandoned. The Fed began purchasing agency MBS in 2008--09 and then long-dated Treasuries through QE1, QE2, the 2011 Maturity Extension Program, QE3, and QE4; by the March 2022 peak the weighted average maturity of SOMA Treasury holdings stood at roughly 8.3 years against 6.2 years for marketable Treasuries as a whole, and the portfolio held approximately \$2.7 trillion of agency MBS with maturities typically over twenty years. Whereas before the GFC nearly all of the SOMA portfolio was held against currency in circulation, today roughly half of it is held against interest-bearing bank reserves and reverse repos: a composition that no longer corresponds to anything Heinsohn and Steiger would recognize as money-against-property.\footnote{Sources on the Fed's maturity structure and composition. On the pre-2008 short-duration SOMA and the post-2008 shift to long-dated Treasuries and agency MBS: Kansas City Fed, ``Considerations for the Longer-Run Maturity Composition of the Federal Reserve's Treasury Portfolio'' (2024): \url{https://www.kansascityfed.org/research/economic-bulletin/considerations-for-the-longer-run-maturity-composition-of-the-federal-reserves-treasury-portfolio/}. On the 8.3-year weighted average maturity figure and the full composition of the post-QE4 portfolio: \citet{GulatiSmith2022}: \url{https://www.kansascityfed.org/Economic\%20Review/documents/9251/EconomicReviewV107N4GulatiSmith.pdf}. Federal Reserve Board's illustrated history of the balance sheet: \url{https://www.federalreserve.gov/econres/notes/feds-notes/a-brief-illustrated-history-of-the-federal-reserves-balance-sheet-20260213.html}. Federal Reserve Board, \emph{Balance Sheet Developments: May 2022}: \url{https://www.federalreserve.gov/monetarypolicy/May-2022-Federal-Reserve-Balance-Sheet-Developments.htm}.}

The Eurosystem transformation is, on the asset-side structural question, even sharper. The pre-2008 ECB operational framework consisted almost entirely of one-week main refinancing operations (MROs) and three-month longer-term refinancing operations (LTROs), with securities held for monetary policy purposes effectively zero. This was the closest any major modern central bank came to the H\&S paradigm: bank-issued money backed by short-dated, private-sector, market-priced claims against commercial bank collateral, rolled at weekly and quarterly frequency. The departure from this paradigm can be traced in steps. In December 2011 and March 2012 the ECB introduced three-year LTROs, extending the maturity of refinancing by more than an order of magnitude over the standard framework. The Targeted LTRO programmes of 2014, 2016, and 2019 (TLTRO I--III) lengthened this further, to four-year maturities with preferential rates. From March 2015 the Public Sector Purchase Programme (PSPP) began outright purchases of euro-area sovereign bonds; from 2016 the Corporate Sector Purchase Programme (CSPP) did the same for investment-grade corporates; the Pandemic Emergency Purchase Programme (PEPP) scaled both from 2020. By end-2021 securities held for monetary policy purposes stood at \texteuro 4.71 trillion, 55\% of total Eurosystem assets; by end-2022 the share had risen to 62.1\% as the refinancing book shrank. The \emph{Handelswechsel}-style refinancing that \emph{Eigentum, Zins und Geld} treats as the institutional paradigm of sound central banking, which the ECB practised, with modifications, until 2008, has effectively disappeared from the asset side.\footnote{Sources on the Eurosystem operational-framework transformation. On the pre-2008 MRO / LTRO framework: Banque de France, ``Refinancing operations'': \url{https://www.banque-france.fr/en/monetary-strategy/operational-framework/refinancing-operations}. On the Eurosystem's current instruments: \url{https://www.ecb.europa.eu/mopo/implement/html/index.en.html}. On the TLTROs introduced in 2014, 2016, and 2019: \url{https://www.ecb.europa.eu/mopo/implement/omo/tltro/html/index.en.html}. On the balance-sheet composition shift 2007--2020: Banque de France, ``Understanding the Expansion of Central Banks' Balance Sheets'': \url{https://www.banque-france.fr/en/publications-and-statistics/publications/understanding-expansion-central-banks-balance-sheets}.}

\subsection{The Tenet and the Option-Value Reading}

Heinsohn and Steiger hold that a sound central bank must maintain a property base sufficient to honour the money it has issued. The position is developed textually in Part D of \emph{Eigentum, Zins und Geld} in connection with the Bundesbank's then-practice of restricting discounting to short-dated bills of exchange, and the authors endorse the practice precisely because of the discipline it imposes on the composition of the asset side \citep[p.~245]{HeinsohnSteiger2004}. Money is created only against property that the central bank has itself encumbered through the issuance; the central bank must therefore hold property it can, in principle, deliver up against presented notes. Sound central banking, on this view, is an exercise in keeping the asset side of the balance sheet in a condition that would permit the institution to meet redemption if redemption were demanded, even under a regime in which it never will be.

Gunnar Heinsohn stated the position sharply and repeatedly in the German-language press after 2008. In a three-part essay in \emph{Die Achse des Guten} in September 2011, written in direct response to the Fed's second round of quantitative easing, he set out the core mechanism. Money is lent only against pledged property and must itself be backed by property of the money-creating bank, he wrote \citep{Heinsohn2011}, and the cost of that encumbrance, the loss of disposition (\emph{Dispositionsverlust}) on the central bank's own property, is what the lending rate compensates. The argument against the unconventional measures is accordingly not about interest rates or liquidity but about the material adequacy of the backing: a central bank that holds paper of doubtful value on its asset side is not, in any operational sense, holding property.

Applied to the post-2008 record, the test does not yield a clean verdict. Technically, all the paper on central-bank asset sides trades in secondary markets and carries a market price: Treasuries, agency MBS, sovereign bonds under PSPP, corporates under CSPP. The test the tenet sets up is nevertheless not satisfied by this observation, for two reasons that deserve to be stated separately. The first is that when the central bank is itself a major buyer, holding 20\% to 40\% of the outstanding stock of a given instrument, as the Fed and Eurosystem have at various points held in Treasuries and in euro-area sovereign bonds respectively, the ``market'' price of the instrument is partly a price the central bank is setting through its own purchases. \citeauthor{GulatiSmith2022}'s estimates put the Fed's SOMA footprint as having lowered the 10-year Treasury yield by more than a percentage point as of 2024:Q2 \citep{GulatiSmith2022}, and QE's declared policy objective during each of its rounds was yield suppression. Whatever one calls this, it is not arm's-length market pricing in the sense that pre-2008 bill-of-exchange refinancing was. The second is that from a consolidated-balance-sheet perspective, central-bank holdings of own-government paper are simply offsetting liabilities: the Treasury's bond becomes the Fed's asset becomes reserves that pay overnight interest, and the composite operation has the same structural signature as direct monetary financing regardless of whether a secondary-market price is stamped on the bond along the way. Whether this counts as debt monetization is contested, and I do not want to overclaim: government bonds carry, at minimum, an implicit claim on the state's taxing power, which is a claim on real economic activity and thereby on property. But the distance between that implicit, second-order backing and the direct first-order property backing that \emph{Eigentum, Zins und Geld} describes is large, and has widened materially since 2008. The asset side has not been replaced entirely by worthless paper; it has been replaced by paper whose backing is real but indirect, whose pricing is partly endogenous to the central bank's own activity, and whose maturity is an order of magnitude longer than the paradigm the authors defended. The tenet has not been respected, but neither has it been grotesquely violated.\footnote{Sources on the yield-suppression effect and the consolidated-balance-sheet perspective. On the Fed SOMA portfolio's estimated effect on the 10-year Treasury yield: \citet{GulatiSmith2022}. On the consolidated-balance-sheet view of SOMA Treasury holdings as offsetting Treasury liabilities: U.S.\ Treasury Borrowing Advisory Committee, \emph{Bill Purchases and the Consolidated Balance Sheet} (Q1 2026 charge): \url{https://home.treasury.gov/system/files/221/TBACCharge1Q12026.pdf}. On the Fed's footprint in the Treasury and agency MBS markets through QE4: NBER Working Paper 30749 (2022): \url{https://www.nber.org/system/files/working_papers/w30749/w30749.pdf}.}

The tenet is normally stated as a prudential requirement. I want to propose reading it instead, or also, as an option. The central bank that holds a well-composed property base against its money issue is maintaining the capacity to revert, if required, to a regime in which redemption is not merely hypothetical but operational. The option is almost never exercised. Under ordinary conditions there is no queue at the window, and the institutional posture of the modern central bank, as distinct from a \emph{Zettelbank}, is precisely that no such queue is permitted to form. But the existence of the option, the fact that the central bank \emph{could}, under sufficiently extreme conditions, revert to a regime with real redemption, constrains behaviour on both sides of the balance sheet and anchors the public's belief in the money's value in something more tangible than the central bank's word. An option need not be exercised to have value. The property-base tenet, in this reading, is the cost of maintaining the option; the discipline on asset-side composition is the option's premium.

The strength of this reading is that it gives the tenet a rationale that a finance theorist recognizes immediately without requiring any commitment to actual redemption. The weakness is that it depends on the option's credibility. An option that all parties know will never be exercised, under any conceivable circumstance, has zero value. Somewhere between ``unexercised but credible'' and ``unexercised and incredible'' there is a threshold below which the option is no longer doing work. Whether the modern central bank's posture is above or below that threshold is an empirical question the paper cannot settle; but the \emph{direction} of motion since 2008 is not in doubt. Balance sheets have become larger and less property-like, redemption as a concept has become more remote, and the distance between the institutional posture and any conceivable reversion has widened materially.

\subsection{The Case for Elimination}

On one reading, the post-2008 record is evidence that the property premium has been institutionally eliminated. The Eurosystem and Federal Reserve have operated at balance-sheet multiples and asset-composition structures that \emph{Eigentum, Zins und Geld} would have classified as insolvent by its own standards, and the predicted consequences: loss of price-level control, currency flight, a crisis of the monetary anchor, have not materialized in any decisive form. Inflation rose sharply in 2021--22 and then receded without a breakdown of the system. Asset prices rose, wealth inequality widened, but these are familiar features of monetary expansion under any theory. The option that Section 6.2 names has been allowed to decay, and the system has continued to function.

On this reading, the Heinsohn/Steiger framework describes one historical institutional configuration for achieving monetary stability, the regime with real redemption, but not a necessary configuration. Fiat monopoly represents a different configuration in which the anchoring of the money's value is achieved by other means: central bank operational independence, inflation targeting, communication policy, the credibility of the policy framework, and ultimately the state's taxing power. Whether or not one admires this architecture, the empirical record of the past decade and a half is that it has not failed. The property-premium mechanism may simply be obsolete: a feature of the \emph{Zettelbank} era that was superseded by institutional developments the authors did not anticipate and would not have endorsed but that have, in fact, held up.

\subsection{The Case for Disguised Operation}

On the other reading, the mechanism has not been eliminated; it has been pushed off the central-bank balance sheet and now operates through other channels. Four are worth naming.

\paragraph{Inflation as continuous-analogue redemption.} In a \emph{Zettelbank} regime the redemption mechanism operates through discrete events: notes presented at the window, property handed over, the bank's balance sheet disciplined ex ante by the threat. Under fiat monopoly the discrete event is gone, but something functionally equivalent remains. When money is issued beyond what the underlying property base can absorb, holders of money do something that resembles redemption: they spend it. Prices rise. The property backing reasserts itself through nominal revaluation rather than through physical handover. The 2021--22 inflation episode, read in this light, was the window finally opening under conditions in which it was no longer possible to keep it closed.

Asset-price inflation is the more focused version of the same channel: money that cannot be redeemed into property through the central bank is redeemed into property directly, with the premium showing up as elevated nominal valuations rather than as bank distress. The S\&P 500 closed 2007 at 1{,}468 and end-2024 at 5{,}881: a price gain of roughly 300\% over seventeen years and, with dividends reinvested, an annualized total return above 10\%. The Euro Stoxx 50, a price-only index, moved from 4{,}399 at end-2007 to 4{,}862 at end-2024: a weaker recovery that in price terms only surpassed its pre-crisis level in late 2021. US residential real estate has more than doubled from its 2012 post-crisis trough, and the Bank for International Settlements (BIS) reports that across advanced economies outside Europe, real residential property prices have risen by roughly 50\% in aggregate since 2010. The euro area recovery has been more modest and more dispersed: the BIS reports an 8\% real increase in euro-area residential property prices since 2010 in aggregate, with wide cross-country variation: Germany and the Netherlands running sharply higher while Italy remains 26\% below its post-GFC level in real terms. These moves proceeded through a sequence of episodes most commentators expected to interrupt them: the 2010--12 European sovereign debt crisis, Brexit, Trump's first term, the pandemic, and the interruptions that did occur (the Q1 2020 sell-off, the 2022 correction) proved small and brief. Gunnar Heinsohn's 2011 essay anticipated this pattern: money that cannot flow into productive lending without viable collateral flows instead into existing property titles, bidding up their nominal prices while underlying economic activity stays flat \citep{Heinsohn2011}. On this reading, that is disguised redemption: the mechanism continues to operate, but at a more diffuse location and slower speed than the \emph{Zettelbank} redemption event.\footnote{Sources on asset-price developments. S\&P 500 daily values: \url{https://fred.stlouisfed.org/series/SP500}. Euro Stoxx 50 annual closing values, compiled from boerse.de via Statista: \url{https://www.statista.com/statistics/261709/} (same series reported in \url{https://en.wikipedia.org/wiki/EURO_STOXX_50}). Case-Shiller U.S.\ National Home Price Index: \url{https://fred.stlouisfed.org/series/CSUSHPINSA}. BIS residential property price statistics, Q4 2024 release: \url{https://www.bis.org/statistics/pp_residential_2505.htm}. Eurostat house price index for the euro area: \url{https://ec.europa.eu/eurostat/statistics-explained/index.php?title=Housing_price_statistics_-_house_price_index}.}

\paragraph{Commercial banking with the mechanism on open display.} At the commercial-bank level the H\&S mechanism operates entirely in view. Deposits are liabilities redeemable on demand against the bank's assets. Redemption of bank liabilities against bank property is precisely what happens in a bank run, and the episodes of early 2023: Silicon Valley Bank, Signature, First Republic, Credit Suisse, are recent demonstrations. What has changed since the \emph{Zettelbank} era is not the mechanism at the bank level but the institutional architecture above it: the central bank has made itself into an unconditional backstop, absorbing the redemption pressure before it can impair the commercial bank's property. The pressure is still priced: into commercial-bank credit spreads, into capital requirements, into deposit-insurance premiums, into the structure of supervisory regulation, but it has been moved up one institutional level and redistributed across the architecture rather than eliminated. This is the aspect of the H\&S framework that practising bankers tend to recognize immediately, which is one reason the theory has always had an audience among practitioners larger than its academic reception would suggest.

\paragraph{The Eurozone sovereign debt crisis.} Between 2011 and 2012 the spreads between German, Italian, Spanish, and Greek sovereign bonds widened sharply, with the differentiation tracking perceptions of each country's fiscal capacity and property base: its capacity to meet its claims with real resources. The ECB had to intervene with Outright Monetary Transactions precisely because the differentiation was threatening to dissolve the currency union. If the property premium were simply absent from the modern system, one would not expect this differentiation to arise at all: all euro sovereigns would trade at the same rate, backed by the same central bank. They did not, and they still do not. Markets continue to price something property-premium-shaped on sovereign debt, and the episode is a live demonstration that the mechanism the H\&S framework describes remains operative at the level of the issuer behind the central bank.

\paragraph{Exchange-rate discipline.} A holder of a fiat currency has no domestic redemption right, but internationally there is an exit: conversion into another currency. A country that over-issues sees its currency depreciate. For reserve currencies the constraint is weak and slow, but for peripheral issuers, for example, the Turkish lira of the past decade, the Argentinian peso repeatedly, any number of historical cases, it has been the binding discipline. The structure is recognizably the same as redemption: holders of money impose a property-premium-type constraint on the issuer by voting with the exit, and the issuer's currency loses value to the extent the property base is perceived inadequate. The mechanism operates across currencies rather than across issuers within a currency, but it is doing the same analytical work.

\medskip

These channels do not restore the property premium to the central-bank balance sheet where Heinsohn and Steiger would have wanted to locate it, and they are weaker than the discrete redemption commitment would be. But they suggest that the mechanism has been displaced rather than eliminated, that the pressure which the option managed is still present in the system, taking more diffuse forms and showing up in different institutional locations.

\section{A Circle Instead of a Conclusion}
\label{sec:circle}

One puzzle that the analysis above leaves unresolved concerns the anchor itself. The book does not raise it, and the position implicit in Heinsohn and Steiger's argument is that property functions as the anchor of the monetary system in virtue of its physical existence and use-value independent of money. That position, on the reading offered here, is harder to sustain than the book treats it as being.

The puzzle is this. In the Heinsohn/Steiger framework the property backing is the non-monetary anchor that grounds the monetary system. But the \emph{nominal magnitude} of the anchor, the number that enters any balance sheet, any pricing calculation, any central-bank prudential assessment, is determined in money. Real estate is priced in money; the bills of exchange and bonds the authors favour as central-bank asset derive their value from discounted cash flows denominated in money; even gold, at any moment in time, trades at a price in money. The reference point that anchors the monetary system is therefore measured in the thing it is supposed to anchor. Pressing on this, the anchor is circular, or, at any rate, the measurement of the anchor is.

A natural response is to say that the circularity is second-order: the anchor has a physical reality independent of its nominal price, and the fact that its numerical representation is endogenous is a measurement artefact rather than a substantive problem. This argument is articulated in \citet[p.~15]{StadermannSteiger2001}. I do not think this response survives contact with the historical record. Bubble episodes: tulip mania, South Sea, 1929, Japanese real estate in the late 1980s, U.S.\ housing 2003--07, cryptocurrencies several times over, are episodes in which the measurement of property value in money goes through self-referential feedback loops that produce nominal magnitudes completely detached from any underlying use-value. The anchor does not hold steady while the money floats around it; the anchor itself becomes a function of monetary dynamics. Bubbles are not anomalies of a property-based monetary system; they are what happens, periodically, when the stabilizing mechanisms that normally dampen the circularity: credit standards, conservative valuation practice, institutional inertia, weaken enough for the latent self-reference to break through. The mechanism driving a bubble is precisely the circularity pressed into a feedback loop: property is worth what it trades for; it trades for what people believe it is worth; beliefs calibrate to observed prices; rising prices justify further belief in rising prices; the loop closes.

Crypto is the sharpest recent instance of the same phenomenon. Bitcoin was designed, on its creator's own statement, as a response to the perceived unmooring of fiat money after 2008: an attempt to re-create something like a commodity-money anchor in digital form.\footnote{Sources for Bitcoin's stated motivation. The white paper: \citet{Nakamoto2008}. The most explicit statement of anti-fiat motivation is Nakamoto's post to the P2P Foundation forum, 11 February 2009, archived at the Satoshi Nakamoto Institute: \url{https://satoshi.nakamotoinstitute.org/posts/p2pfoundation/1/}: ``The root problem with conventional currency is all the trust that's required to make it work. The central bank must be trusted not to debase the currency, but the history of fiat currencies is full of breaches of that trust.'' The Bitcoin genesis block (block 0, mined 3 January 2009) embeds in its coinbase parameter the headline ``The Times 03/Jan/2009 Chancellor on brink of second bailout for banks,'' a verbatim reference to the UK's response to the 2008 financial crisis; see \url{https://en.bitcoin.it/wiki/Genesis_block}.} With ordinary property the anchor at least exists outside the monetary system, even if its nominal measurement is circular. With Bitcoin there is no external referent at all. The issuance is on no balance sheet's passive side. The only anchor is a 21-million cap that is a rule inside the protocol. The scarcity being relied on and the system relying on it are the same object. And the asset is denominated and traded in fiat: its value is what someone will pay in dollars or euros for it, where those currencies are themselves only partially anchored in property after 2008. The circularity is complete. One cannot get a cleaner instance. The extreme volatility of crypto prices is what one would predict for a monetary object whose stabilizing anchor mechanism has been dispensed with. The stablecoin phenomenon (USDC, USDT backed by Treasuries) is a partial re-import of the property structure, the crypto ecosystem's implicit admission that pure protocol-scarcity is inadequate as a monetary anchor, and an acknowledgement, in its own vocabulary, that the Heinsohn/Steiger mechanism is what was missing.

Where this leaves the question Section~\ref{sec:after2008} opened on, I am not sure. The case for elimination (\S6.3) and the case for disguised operation (\S6.4) are framed as alternative readings of a common record, and the weight of the evidence in \S6.4 persuades me that the mechanism has not been eliminated. that it operates in displaced form through channels the central-bank balance sheet no longer shows. But the puzzle of this section sits underneath both readings. If the property anchor is itself only semi-stable, subject to periodic circularity episodes in which its nominal magnitude detaches from its material substance, then neither the \emph{Zettelbank} regime nor the fiat monopoly regime has access to the external reference point that would, in the ideal, ground monetary valuation non-circularly. The two regimes differ in how they \emph{manage} the circularity: the \emph{Zettelbank} regime with discrete redemption points as periodic reality-checks, the fiat monopoly regime with diffuse and slower checks in the form of inflation, asset-price adjustment, exchange-rate drift, sovereign-spread differentiation. But neither eliminates it. What Heinsohn and Steiger gave the literature, on this reading, is the clearest available description of the institutional configuration that manages the circularity best, and the reasons it does so. What they would have said about the configurations that, having dispensed with that management, have nevertheless not yet failed remains an open question.

\appendix
\section{Mathematical Appendix}
\label{app:math}

\noindent\emph{The derivations below provide the formal apparatus for the claims in Sections 2--5. They are developed following \citet{Varian2010} for ease of exposition. \citet[chap.~3]{MWG1995} could be deployed for a more condensed treatment.}

\subsection{Slutsky Decomposition for the Intertemporal Choice}
\label{app:slutsky}

\paragraph{Setup.} An agent lives for two periods, with income endowment $(m_1, m_2)$ and a single consumption good in each period. Lending and borrowing are available at the gross interest rate $1+r$. Let $(c_1, c_2)$ denote consumption. The intertemporal budget constraint, written in present-value form, is
\[
c_1 + \frac{c_2}{1+r} \;=\; m_1 + \frac{m_2}{1+r}.
\]
Preferences are time-separable:
\[
U(c_1, c_2) \;=\; u(c_1) + \beta\, u(c_2), \qquad \beta \;=\; \frac{1}{1+\rho},
\]
with $u$ strictly increasing and strictly concave and $\rho \geq 0$ the pure rate of time preference. The agent's savings in period 1 are $s = m_1 - c_1$.

\paragraph{The relative price of present consumption.} The interest rate $r$ is the relative price of $c_1$ in units of $c_2$: to obtain one extra unit of future consumption the agent must give up $1/(1+r)$ units of present consumption. An increase in $r$ raises the price of $c_1$ relative to $c_2$.

\paragraph{Slutsky decomposition.} The effect of a change in $r$ on optimal $c_1$ decomposes into a substitution and an income effect. For a consumer with an endowment \citep[chap.~9]{Varian2010}, the appropriate version of the Slutsky equation is
\[
\frac{\partial c_1}{\partial r} \;=\; \underbrace{\left.\frac{\partial c_1}{\partial r}\right|_{U}}_{\text{substitution effect}} \;+\; (m_1 - c_1)\,\underbrace{\frac{\partial c_1}{\partial W}}_{\text{income effect}}
\]
where $W$ is lifetime wealth. The substitution term, the change in $c_1$ along the indifference curve as the relative price of $c_1$ rises, is unambiguously negative: present consumption falls and savings rise. The income term carries the factor $(m_1 - c_1) = s$, the agent's net trading position in period 1. For a \emph{net saver} ($s > 0$) a higher interest rate raises lifetime wealth; if $c_1$ is a normal good the income effect raises $c_1$ and \emph{reduces} savings, opposing the substitution effect. For a \emph{net borrower} ($s < 0$) the income effect lowers $c_1$ and reinforces the substitution effect.

\paragraph{Sign of the net effect on savings.} Writing $s = m_1 - c_1$ and applying the decomposition:
\begin{itemize}
\item For a borrower, $\partial s / \partial r > 0$ unambiguously.
\item For a saver, $\partial s / \partial r$ may be positive, negative, or zero. The two effects can cancel.
\end{itemize}
This is the formal statement of the ambiguity. The direction of the response of aggregate savings to the interest rate is an empirical question that depends on preferences (in particular the intertemporal elasticity of substitution) and on the distribution of savers and borrowers in the population. There is no clean theoretical prediction.

\paragraph{A cancellation example.} With logarithmic utility $u(c) = \ln c$, for which the intertemporal elasticity of substitution equals one, and a saver whose entire endowment is in period 1 ($m_2 = 0$), the optimum can be computed in closed form. Maximizing $\ln c_1 + \beta \ln c_2$ subject to $c_1 + c_2/(1+r) = m_1$ gives the first-order condition
\[
\frac{1}{c_1} \;=\; \beta\,(1+r)\,\frac{1}{c_2}\,\frac{1}{1+r} \;=\; \frac{\beta}{c_2},
\]
so $c_2 = \beta\,(1+r)\,c_1$. Substituting into the budget constraint yields
\[
c_1 \;=\; \frac{m_1}{1+\beta} \;=\; \frac{m_1}{1 + \frac{1}{1+\rho}}.
\]
Period-1 consumption, and therefore savings, depends on $\rho$ but not on $r$. The interest elasticity of savings is exactly zero. The level of $\rho$ can be set anywhere in $[0, \infty)$ without affecting this result: zero elasticity is consistent with any rate of time preference, including very high ones. This refutes the inference from zero elasticity to no time preference.

\subsection{Two-Stage Portfolio Problem and the Euler Equation}
\label{app:euler}

\paragraph{Setup.} Extend the model of \S\ref{app:slutsky} by allowing the agent to save in two distinct stores of value: cash $M$, which yields no interest, and bonds $B$, which pay the gross return $1 + r$ in period 2. In the simplest form (no aggregate uncertainty, a single deterministic interest rate) the problem is
\[
\max_{c_1, c_2, M, B}\; u(c_1) + \beta\, u(c_2)
\]
subject to
\[
c_1 + M + B \;=\; m_1, \qquad c_2 \;=\; m_2 + M + (1+r)\,B, \qquad M \geq 0,\; B \geq 0.
\]

\paragraph{First-order conditions.} Let $\mu \geq 0$ and $\nu \geq 0$ be the multipliers on $M \geq 0$ and $B \geq 0$. After substituting out the multiplier on the period-1 budget constraint, the first-order conditions with respect to $M$ and $B$ read:
\[
u'(c_1) \;\geq\; \beta\, u'(c_2), \qquad \text{equality if } M > 0,
\]
\[
u'(c_1) \;\geq\; \beta\,(1+r)\, u'(c_2), \qquad \text{equality if } B > 0.
\]

\paragraph{Cash is dominated for $r > 0$.} If both $M > 0$ and $B > 0$ held simultaneously, the two equalities would imply $\beta\, u'(c_2) = \beta\,(1+r)\,u'(c_2)$, hence $r = 0$. For any $r > 0$, therefore, $M = 0$ at the optimum and $B > 0$: the agent saves entirely in bonds.

\paragraph{The resulting Euler equation.} Setting $M = 0$ and using the bond condition with equality:
\[
u'(c_1) \;=\; \beta\,(1+r)\, u'(c_2) \;=\; \frac{1+r}{1+\rho}\, u'(c_2).
\]
This is the standard intertemporal Euler equation. It governs the consumption-saving choice, that is, how much the agent saves, given the interest rate $r$. Time preference $\rho$ appears explicitly inside the condition. Keynes's observation that interest is paid only when liquidity is given up is correctly captured (cash earns nothing, bonds earn $r$, so saving in cash is dominated whenever $r > 0$). The observation does not eliminate $\rho$; it relocates the market in which the interest rate $r$ is determined, leaving the preference parameter that governs how much the agent chooses to save at that $r$ untouched.

\paragraph{A stationary case.} If consumption is stationary ($c_1 = c_2 = c$) the Euler equation simplifies to $1 = \beta(1+r)$, that is
\[
1 + r \;=\; \frac{1}{\beta} \;=\; 1 + \rho, \qquad \text{i.e.,}\quad r \;=\; \rho.
\]
The risk-free rate of interest in a stationary equilibrium equals the rate of time preference. With $\rho = 0$ the equilibrium risk-free rate is zero. Time preference is precisely what generates a positive risk-free rate in a world with stable consumption: the foundational fact that the Heinsohn/Steiger framework, having dispensed with $\rho$, has no resources to explain.

\subsection{Risk Premium on a Defaultable Bond}
\label{app:risk}

\paragraph{Setup.} A lender can hold either of two one-period bonds. The risk-free bond pays $1 + r_f$ with certainty. The risky bond promises gross return $1 + \tilde{r}$, but the borrower defaults with probability $\pi \in [0,1]$. In default the lender recovers only a fraction $1 - \lambda$ of the promised payment, where $\lambda \in [0,1]$ is the loss given default.

\paragraph{Risk-neutral pricing.} Indifference between the two bonds requires equality of expected gross returns:
\[
(1 - \pi)(1 + \tilde{r}) \;+\; \pi\,(1 + \tilde{r})(1 - \lambda) \;=\; 1 + r_f.
\]
Factoring $1 + \tilde{r}$ on the left-hand side,
\[
(1 + \tilde{r})(1 - \pi\lambda) \;=\; 1 + r_f,
\]
so that
\[
1 + \tilde{r} \;=\; \frac{1 + r_f}{1 - \pi\lambda}.
\]
For $\pi\lambda$ small, the typical case for performing collateralized credit, a first-order Taylor expansion gives
\[
\tilde{r} \;\approx\; r_f + \pi\lambda.
\]
The spread $\pi\lambda$ is the expected per-unit loss from default.

\paragraph{Risk-averse pricing.} A risk-averse lender whose marginal utility covaries with the bond's payoff requires an additional compensation for variance. In a standard consumption-based asset-pricing framework \citep{Lucas1978,Breeden1979,Cochrane2005} the premium is augmented by a term proportional to the covariance between the bond's payoff and the lender's intertemporal marginal rate of substitution. Writing the full risk premium as
\[
\tilde{r} - r_f \;=\; \pi\lambda \;+\; \text{(variance-related term)}
\]
accommodates both the risk-neutral and risk-averse cases at the level of generality required in the main text.

\paragraph{Connection to A.2.} Combining with the stationary-equilibrium result of Appendix~\ref{app:euler},
\[
r_f \;=\; \rho,
\]
the full two-term decomposition referred to in Section~\ref{sec:decomp} is
\[
\tilde{r} \;=\; \underbrace{\rho}_{\substack{\text{time preference}\\(\ref{app:euler})}} \;+\; \underbrace{(\tilde{r} - r_f)}_{\substack{\text{risk premium}\\(\ref{app:risk})}}.
\]
The first term is the price of time; the second is the price of risk. Both are present in ordinary collateralized lending, and the property premium in its general form is identified in the main text with the second. The claim examined in Section~\ref{sec:bank} is that when the lender is a money-issuing bank with real redemption obligations a third term enters the decomposition: a term that neither \ref{app:euler} nor \ref{app:risk} captures and that standard asset pricing does not articulate.

\subsection{The Money-Issuing Bank and the Third Term}
\label{app:bank}

\paragraph{Setup.} A bank makes a one-period loan of size $L$ to a borrower whose default probability is $\pi$ and loss given default $\lambda$, as in \S\ref{app:risk}. Unlike the lender of \S\ref{app:risk}, the bank funds the loan by issuing its own circulating notes $N = L$ redeemable on demand against property held by the bank. The notes may be presented by any holder at any date $t \in [0,1]$ during the life of the loan. Let $R$ denote the amount of liquid property, gold, other banks' notes, central-bank-eligible collateral, the bank holds as redemption reserve over the life of the loan, and let $\kappa$ denote the opportunity cost per unit and per period of holding property in that form rather than in its next-best unencumbered use. The bank's redemption reserve is chosen to satisfy its redemption obligation with acceptable probability:
\[
R \;=\; r^\ast(N) \;=\; r^\ast(L),
\]
where $r^\ast(\cdot)$ is a required-reserve function determined by the distribution of redemption demand and the bank's tolerance for redemption failure. We take $r^\ast$ to be increasing in $L$ and strictly positive for $L > 0$; the first property says more outstanding notes require more reserves, the second that \emph{any} positive note issue imposes \emph{some} redemption exposure.

\paragraph{The bank's per-unit cost of lending.} The bank's per-unit cost of making the loan has three components: the risk-free opportunity cost of the funds, the expected default loss on the loan, and the opportunity cost of the associated redemption reserve. Writing them in order,
\[
\text{cost per unit lent} \;=\; r_f \;+\; \pi\lambda \;+\; \kappa\,\frac{r^\ast(L)}{L}.
\]
Competitive pricing sets the bank's lending rate equal to its per-unit cost. Identifying the third term as
\[
\phi \;\equiv\; \kappa\,\frac{r^\ast(L)}{L},
\]
the three-term decomposition of the main text is recovered:
\[
\tilde{r}_{\text{Zettelbank}} \;=\; r_f \;+\; \pi\lambda \;+\; \phi.
\]

\paragraph{Two properties of $\phi$.} The construction makes clear why $\phi$ cannot be absorbed into either of the other two terms. First, $\phi$ is present even when the loan is perfectly safe: $\pi = 0$ implies $\pi\lambda = 0$, but $r^\ast(L) > 0$ and therefore $\phi > 0$. The third term is orthogonal to default risk. Second, $\phi$ is independent of intertemporal substitution: it does not depend on the bank's preferences over the timing of its own consumption and does not involve the discount factor $\beta$. It is a pure opportunity cost attached to the bank's balance sheet, distinct from both $r_f$ and from the risk premium of \S\ref{app:risk}.

\paragraph{Limiting cases.} Two limiting cases recover the results stated in the main text:
\begin{itemize}
\item \emph{No money creation} ($N = 0$): the lender of \S\ref{app:risk}, not a bank in the technical sense used here. Then $r^\ast(0) = 0$ and $\phi = 0$; the two-term decomposition of Section~\ref{sec:decomp} obtains.
\item \emph{No real redemption} ($\kappa = 0$): a central bank with a monopoly on issuance and no redemption obligation. The opportunity cost of holding redemption reserves vanishes; $\phi = 0$ regardless of $r^\ast$; the two-term decomposition again obtains.
\end{itemize}
The third term is therefore strictly tied to the institutional configuration of competitive issuance with a binding redemption obligation. The property premium in the specific and original sense, Heinsohn and Steiger's genuine contribution to the theory of interest, is the component of the bank's lending rate associated with this configuration, and with it alone.

\bibliographystyle{plainnat}
\bibliography{references}

\end{document}